\documentclass[aip,preprint]{revtex4-2}
\usepackage{graphicx}
\usepackage{amsmath}
\usepackage{booktabs}
\usepackage{float}

\begin{document}


\title{Unified Theory of Dark Count Rate and System Detection Efficiency for NbN, WSi Based Superconducting Single Photon Detectors} 



\author{Daien He}\altaffiliation{These authors contributed equally to the work}
\author{Leif Bauer}\altaffiliation{These authors contributed equally to the work}
\author{Sathwik Bharadwaj}
\affiliation{The Elmore Family School of Electrical and Computer Engineering, Purdue University, West Lafayette, 47907, IN, USA}
\author{Zubin Jacob}\email{zjacob@purdue.edu}
\affiliation{The Elmore Family School of Electrical and Computer Engineering, Purdue University, West Lafayette, 47907, IN, USA}


\date{\today}


\begin{abstract}
Predicting the behavior of superconducting nanowire single photon detectors (SNSPDs) is important as their use becomes more widespread in fields ranging from quantum computing to quantum remote sensing. Here, we present a vortex crossing theory of photon detection which provides a unified definition of system detection efficiency and dark count rates. Our approach quantitatively captures the plateau region of system detection efficiency for NbN and WSi based SNSPDs. We concurrently predict the temperature dependence of dark count rates and the intrinsic timing jitter of SNSPDs. We extensively benchmark our model against various experiments to aid in the design of the next generation of SNSPDs.
\end{abstract}

\maketitle 

\section{Introduction}

Superconducting nanowire single photon detectors (SNSPDs) are notable for their high sensitivity across a wide range of wavelengths \cite{reddy_superconducting_2020,marsili_detecting_2013,verma_single-photon_2021}, low dark count rates, and precise timing resolution \cite{korzh_demonstration_2020}. Due to their extraordinary capabilities, SNSPDs have led to innovations in quantum science \cite{shalm_strong_2015,you_superconducting_2020,alexander_manufacturable_2025}, LIDAR \cite{esmaeil_zadeh_superconducting_2021,bao_photon_2024}, deep space communications \cite{wollman_snspd-based_2024,hao_compact_2024}, biological imaging \cite{tamimi_deep_2024,buschmann_integration_2023}, and quantum communications \cite{you_superconducting_2020}. Despite this success, SNSPD modeling has remained a difficult process. Many promising designs utilize amorphous\cite{,marsili_detecting_2013} or polycrystalline films \cite{zhang_nbn_snspd}, whose properties are difficult to predict from ab-initio methods\cite{du_atomistic_2022}. Additionally, relevant material parameters often depend on film deposition, substrate, and fabrication \cite{luo_niobium_2023}. Only recently has rapid progress occurred on the theoretical mechanism underlying the detection process  \cite{engel_detection_2015,jahani_probabilistic_2020,simon_ab_2025,allmaras_modeling_2020}.

Models of the SNSPD detection mechanism have primarily focused on various versions of vortex theory \cite{jahani_probabilistic_2020,engel_detection_2015} or microscopic phase slip theory \cite{simon_ab_2025,allmaras_modeling_2020}. Several vortex models have successfully predicted experimental device metrics. Engel, et. al\cite{engel_detection_2015} proposed a vortex model which successfully predicted system detection efficiency (SDE) in an NbN experiment from position dependent effects. A probabilistic model proposed by Jahani, et. al\cite{jahani_probabilistic_2020} captured the dependence of timing jitter on photon energy. Phase slip models have also shown some success in matching with experiments. A phase slip model proposed by Allmaras, et. al\cite{allmaras_modeling_2020} captured the dependence of timing jitter on photon energy. Another phase slip model proposed by Simon, et. al\cite{simon_ab_2025} found close agreement with the SDE plateau region over a range of widths, temperatures, and wavelengths. However, simultaneous prediction of SDE and dark count rate has proven challenging. Additionally, existing SNSPD detection models have not yet demonstrated comprehensive benchmarking against experiments across different material systems.

In this paper we will focus on applying the vortex crossing model to NbN and WSi material systems which have widely reported material properties. We will demonstrate that our model can capture the temperature dependent dark count rate versus bias current. Additionally, we will show that the plateau region of SDE can be predicted using material parameters extracted from experiments. Lastly, we will present a model for the SNSPD timing jitter which combines geometric and vortex jitter. Our work can aid in the search for new SNSPD materials as well as new device geometries which optimize the extraction of information from every incoming photon.

\section{Vortex Dynamics in SNSPDs}

\subsection{Detection Mechanism based on Vortex Crossing}
Vortices are a local destruction of the order parameter with circulating current containing magnetic flux of $\Phi_0 = h/2e$ \cite{tinkham_introduction_2015}. Ginzburg-Landau theory tells us that the singular nature of the complex superconducting order parameter $\psi=|\psi|e^{i\varphi}$ causes vortex flux to be quantized \cite{tinkham_introduction_2015}. At locations where the order parameter is reduced to zero, a jump discontinuity occurs in the phase. This discontinuity causes the circulation of current around that point to occur in discrete values, proportional to the magnetic flux quantum $\Phi_0$ \cite{tinkham_introduction_2015}. Since flux contained in superconductors is quantized, vortices are the smallest stable destruction of the superconducting state. 2D XY-type models of superconducting films also present vortices as the preferred excitation \cite{minnhagen_two-dimensional_1987}. These vortices can nucleate from nearby magnetic fields \cite{tinkham_introduction_2015}, latent thermal energy \cite{minnhagen_two-dimensional_1987}, or thermal excitations \cite{zotova_photon_2012}. Additionally, vortex nucleation rates can be enhanced when the nanowire edge contains defects \cite{benfenati_vortex_2020}.

\begin{figure*}[ht!]
\centering
\includegraphics[width=\textwidth]{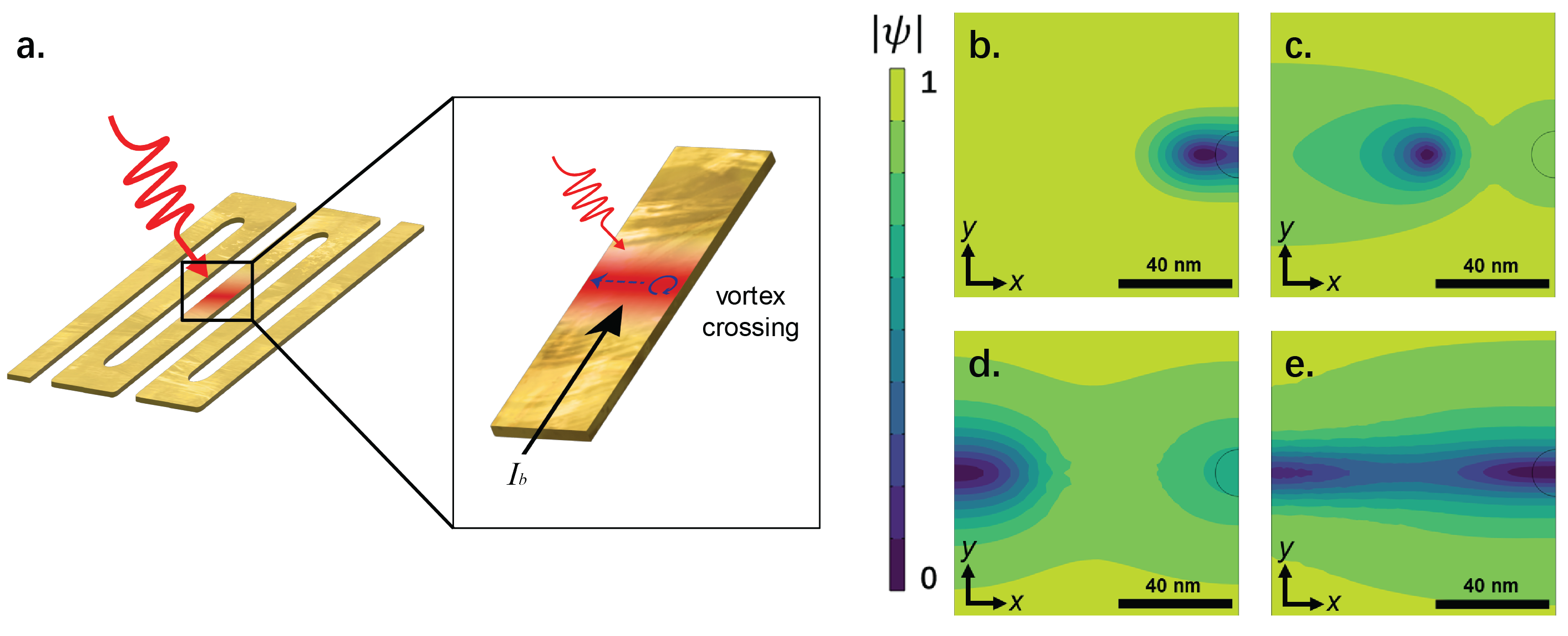}
\caption{Basic mechanics of vortex crossing phenomenon in superconducting nanowire single-photon detector (SNSPD). (a) When the current-carrying nanowire absorbs a photon, it breaks the Cooper pairs into quasi-particles, thus disrupting the superconducting state locally. The superconducting current is redistributed, reducing the potential barrier that limits the vortex movement. Due to the force exerted by the bias current, this vortex will move across the nanowire. (b-e) Superconducting order parameter simulation based on time-dependent Ginzburg-Landau theory. (b) A vortex core with $|\psi|=0$ is nucleated from the edge due to thermal energy from the incident photon. (c) \& (d) With a current bias in the $+y$ direction, the vortex moves toward the other edge and disrupts superconductivity along the path. (e) This vortex crossing path eventually turns into a normal-conductivity `belt' with a suppressed order parameter. }
\label{Fig:vortex_crossing}
\vspace{-10pt}
\end{figure*}

The motion of a vortex is limited by a potential barrier inside the nanowire \cite{PhysRevB.83.144526,engel_detection_2015}. The current-induced Magnus force works against this potential to push the vortex in a direction perpendicular to the current\cite{jahani_probabilistic_2020,PhysRevB.77.174517}. As the vortex moves across the width of the nanowire it disturbs the superconducting phase causing a 2$\pi$ phase difference perpendicular to its path \cite{PhysRevB.77.174517}. Once the vortex crosses the nanowire, a 2$\pi$ phase difference has been generated across the entire width. This causes a normal belt to form across the path, leading to a phase transition in the SNSPD (see Fig. \ref{Fig:vortex_crossing}a)\cite{PhysRevB.77.174517}. This phase transition caused by vortex motion is one of the mechanisms responsible for the `clicking' events observed in SNSPD voltage \cite{engel_numerical_2013, epstein_vortex-antivortex_1981, PhysRev.164.498, PhysRevB.77.174517}.

To gain deeper insight into the detection mechanism governed by vortex dynamics in superconducting materials, we utilize the time-dependent Ginzburg–Landau (TDGL) theory to investigate the behavior of vortices in SNSPDs. Specifically, we develop a 2D TDGL model to simulate vortex dynamics in a current-biased superconducting nanowire structure. We use a diffusive hotspot to simulate the effects of photon incidence or thermal fluctuations. This TDGL modeling approach enables a qualitative evaluation of the vortex crossing process. As illustrated in Fig. \ref{Fig:vortex_crossing}(b-e), we simulate a vortex nucleating at the edge of the nanowire immediately following photon absorption. The order parameter $\psi$ is suppressed at the vortex core, marking the initial disruption of superconductivity. As the vortex moves across the nanowire due to the bias current in the $+y$ direction, it disturbs the local superconducting phase. This vortex-induced 2D phase-slip results in a belt-like suppression of superconductivity across the nanowire.

Vortex crossing events can also be caused by thermal fluctuations from the latent thermal energy in a superconductor \cite{PhysRevB.77.174517,engel_detection_2015}. Therefore, the combined effect of the temperature, bias current, and barrier determine the probability of a vortex crossing event. A previous work \cite{jahani_probabilistic_2020} proposed a probabilistic criterion for the detection mechanism based on a single vortex crossing at a finite temperature. This vortex crossing model showed that the bias current redistribution caused by photon absorption can significantly increase the probability of vortex crossing, even if the barrier has not completely vanished.  In this paper we present an adapted version of this approach using thermal activation theory and temperature dependent models of material parameters.

This paper will focus on a vortex crossing theory of detection. However, our unified theory can be extended to other types of detection events such as phase slips or vortex-antivortex unbinding. This extension begins with determining the energy barrier of a given type of detection event. The energy barrier can be obtained from time-dependent Ginzburg-Landau theory by applying the string method\cite{e_string_2002} to determine the minimum energy path\cite{PhysRevB.85.174507}. Then, the energy barrier can be computed by finding the free energy at each point along the path \cite{PhysRevB.85.174507}. We have found that the free energy barrier for vortex crossing from TDGL matches London theory with a constant offset (see Supplementary Material). For phase slips the minimum energy path involves changes to the configurations of order parameter in two-dimensional space\cite{PhysRevB.77.174517}. This can be computed by extending a similar string method which involves parameterizing the changes in order parameter configurations \cite{benfenati_vortex_2020}. We are not predisposed towards a particular detection mechanism, and there may be some geometries or materials where other detection mechanisms are more probable \cite{ZHANG2022100006}. In future works we plan to extend our model to these other types of detection events. 

\subsection{Vortex Barrier Energy Model}
 We perform numerical calculations to simulate the detection event involving the following steps: generation and diffusion of hot electrons after photon incidence, breakdown of Cooper pairs into quasi-particles, and consequently, redistribution of the bias current. These events lower the potential barrier $U_{max}$ faced by a single vortex to cross the width of the SNSPD, which significantly increases the vortex crossing probability \cite{jahani_probabilistic_2020}. Using the quantified vortex crossing rate, we can compute important characteristics of the SNSPD device physics, such as dark count rate and system detection efficiency.

We use finite-difference methods to simulate the diffusion processes of hot electrons and quasi-particles, enabling us to calculate the inhomogeneous current in the nanowire. The diffusion processes for hot electrons and quasi-particles under the photon energy $h\nu$ are given by:\cite{engel_detection_2015}
 \begin{equation}
     \frac{\partial C_e(\mathbf{r}, t)}{\partial t} = D_e\nabla^2C_e(\mathbf{r}, t) 
     \label{eq:quasi_1}
 \end{equation}
 \begin{widetext}
 \begin{equation}
     \frac{\partial C_{qp}(\mathbf{r}, t)}{\partial t} = D_{qp}\nabla^2C_{qp}(\mathbf{r}, t) - \frac{C_{qp}(\mathbf{r}, t)}{\tau_r} + \frac{\varsigma h\nu}{\Delta\tau_{qp}}\left(\frac{n_{se,0}-C_{qp}(\mathbf{r}, t)}{n_{se,0}}\right)e^{-t/\tau_{qp}}C_e(\mathbf{r}, t)
     \label{eq:quasi_2}
 \end{equation}
 \end{widetext}
where $C_e(\mathbf{r}, t)$ denotes the hot electron distribution density, and $C_{qp}(\mathbf{r}, t)$ is the quasi-particle distribution density. Here $D_e$, $D_{qp}$, $\tau_r$, $\tau_{qp}$, and $n_{se,0}$ are the hot-electron diffusion coefficient, quasi-particle diffusion coefficient, recombination time, lifetime of quasi-particles multiplication process, and density of superconducting electrons before the photon absorption, respectively. $\varsigma$ is the quasi-particle multiplication efficiency, which has been assumed constant ($\sim 25\%$). The current distribution can be found from the following equation:
\begin{equation}
    \nabla\cdot\mathbf{j}(\mathbf{r}, t) = \nabla\cdot\left(\frac{\hbar}{m}n_{se}(\mathbf{r}, t)\nabla\varphi(\mathbf{r}, t)\right) = 0
\end{equation}
 \begin{equation}
    n_{se}=n_{se,0}-C_{qp}
\end{equation}
where $n_{se}(\mathbf{r}, t)$ represents the density of superconducting electrons after the photon absorption, $\varphi$ is the phase of the superconducting order parameter, $\hbar$ is the reduced Planck's constant, and $m$ is the mass of an electron.

In the time-dependent Ginzburg–Landau (TDGL) theory, the barrier $U_{max}$ is given by the difference in free energy\cite{PhysRevB.85.174507} as
\begin{equation}
     U_{max}=F_{saddle}-F_{ground}-\frac{\hbar}{2e}\frac{I}{I_c}\Delta\varphi
 \end{equation}
where $F_{saddle}$ is the GL free energy of the saddle point state (a metastable state between two local minima) \cite{PhysRevB.85.174507, PhysRevB.77.174517, PhysRevB.83.144526, PhysRevLett.83.2409} and $F_{ground}$ is the free energy of the initial superconducting state. $e$ is the electron charge, $I$ is the applied current, $I_c$ is the critical current, and $\Delta\varphi$ represents the difference in phase of order parameter $\psi$ between the current-in port and current-out port. The GL free energy in its dimensionless form is calculated based on the equation \cite{PriourDefect}
\begin{eqnarray}
     F_{GL}=\nonumber&&\int\left[\left|\psi^*(\vec{\nabla}/i-\vec{A})\psi\right|^2-|\psi|^2\right.\\&&\left.+\frac{\kappa^2}{2}|\psi|^4+\left|\vec{\nabla}\times\vec{A}\right|^2\right]d^3x\\\nonumber
 \end{eqnarray}
where $\vec{A}$ is the magnetic vector potential, and $\kappa$ is the ratio of London penetration depth to superconducting coherence length. The order parameter $\psi$ is calculated using TDGL equations (see Supplementary Material).

Since the barriers obtained from TDGL matches London theory, we adopt the London-based quasiparticle diffusion approach to estimating the barrier as shown below:
\begin{eqnarray}
\frac{U_{max}}{\epsilon_0}= \nonumber&&\max_{x_\nu}\left[\frac{\pi}{W}\int^{x_v'}_{\frac{\xi-W}{2}}
\frac{n_{se}(x',t)}{n_{se,0}}\tan\left(\frac{\pi x'}{W}\right)dx'\right.\\
&&-\frac{2W}{I_c\exp(1)\xi}\int^{x_v'}_{\frac{-W}{2}}
\left.\frac{n_{se}(x',t)}{n_{se,0}}j_y(x',t)dx'\right]\\\nonumber
\end{eqnarray}
where $W$, $d$, $x_v'$, $\xi$, $j$  are the nanowire width, nanowire thickness,  vortex position, coherence length, and current density respectively. The vortex barrier energy is normalized to the characteristic vortex energy $\epsilon_0=\frac{\Phi_0^2d}{4\pi\mu_0\lambda^2}$, where $\Phi_0$ is the magnetic flux quantum, $\mu_0$ is the vacuum permeability, and $\lambda$ is the magnetic penetration depth.

\section{Unified theory of dark count and SDE}
By incorporating the barrier energies above into our vortex crossing model, we can predict device performance such as dark count rate and system detection efficiency through the vortex crossing rate $\Gamma_{v}$. From thermal activation theory\cite{hanggi_reaction-rate_1990}, we can calculate the vortex crossing rate using
\begin{equation}
    \Gamma_v = \alpha_v\exp(-U_{max}/k_BT)
\end{equation}
\begin{equation}
\alpha_v(T)=\alpha_0\exp(-\beta_1/k_BT)
\end{equation}
where $k_B$ is Boltzmann's constant, $T$ is temperature, and $\alpha_v(T)$ is the vortex attempt rate. Here $\alpha_0$ represents the zero barrier vortex attempt rate and $\beta_1$ is the barrier to vortex nucleation\cite{minnhagen_two-dimensional_1987} with quantum corrections\cite{PhysRevA.41.5366}. We define the dark count rate (DCR) of the system as
\begin{equation}
    DCR = \Gamma_v
\end{equation}
where $C_e = 0$, $C_{qp} = 0$,  $n_{se}(\mathbf{r}, t) = n_{se,0}$, $j(\mathbf{r}, t) = j_0$, where $j_0$ is the current distribution prior to photon incidence. The attempt rate $\alpha_v$ refers to how often a vortex tries to cross the free-energy barrier in the superconducting nanowire. In a bulk superconductor, the vortex attempt rate can be measured directly through the frequency response of the critical current\cite{FABREGA19942981}. For thin films, one can evaluate the attempt rate through the effective film sheet resistance by treating the vortex as a particle\cite{bulaevskii_vortex-assisted_2012}. A more direct approach to determine vortex attempt rate is to fit this thermally-activated model to experimentally measured dark count rates\cite{bartolf_current-assisted_2010} in SNSPDs. The parameters $\alpha_0$ and $\beta_1$ are modeled to fit the measured dark counts under a specific temperature. When measurements are available across multiple temperatures, we compute the mean values of $\alpha_0$ and $\beta_1$ and use these to define $\alpha_v(T)$ via the best-fit model to the full set of dark-count data. Following the dark-count modeling approach, our study indicates that the parameters $\alpha_0$ and $\beta_1$ are inherently material-dependent. Consequently, by determining these parameters from experimental data for a specific superconducting material, they can be utilized within our model to facilitate further investigations of the behavior of the same material in SNSPD applications. 

The vortex crossing rate $\Gamma_{v}$ can furthermore be used to predict the plateau region in the current-dependent system detection efficiency (SDE). Following Jahani, et. al\cite{jahani_probabilistic_2020}, we model the detector response as a Poisson process with rate equal to $\Gamma_v$. Incident light will cause $\Gamma_v$ to increase due to the reduction in vortex crossing barrier from the photon-induced hot-spot. In a Poisson process with a constant count rate $\Gamma$, the probability that at least one count occurs in time $t$ is given by $P(k\geq1) =1-\exp(-\Gamma t)$. To account for the accumulated effects of the time varying count rate, we take the integral of $\Gamma_v(t')dt'$ over $t$. Therefore, we define the SDE as
\begin{equation}
    SDE=1-\exp\left(-\int^t_0\Gamma_{v}(t')dt'\right)
\end{equation}
This general framework unifies photon counts and dark counts: a defect is nucleated in both cases, but only thermal fluctuations produce remnant dark counts in the absence of light. The energy barrier $U_{max}$ thus governs both the system’s photon detection efficiency and its dark-count rate.

\section{Extensive benchmarking against experiments}

\subsection{NbN based SNSPD}

\begin{figure*}[ht!]
\centering
\includegraphics[width=\textwidth]{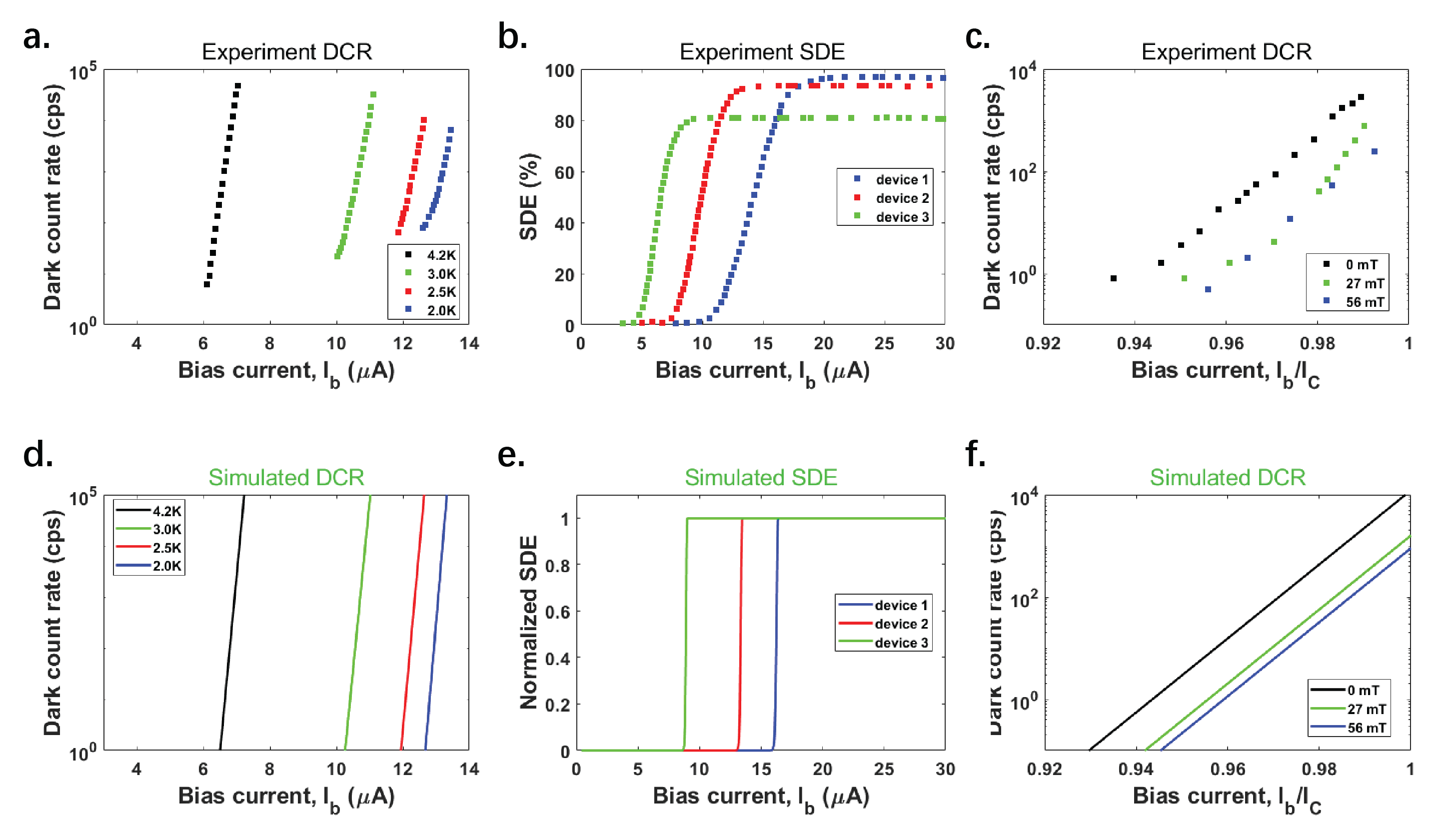}
\caption{NbN based SNSPD Experiment Vs. Simulation. The first row (a-c) shows the experimental results of measured dark count rate (DCR) and system detection efficiency (SDE) retrieved from (a)\&(b) Zhang's work \cite{zhang_nbn_snspd}, and (c) Charaev's work \cite{charaev_snspd}. The second row (d-f) shows the simulation results of our vortex crossing model based on the material parameters and test conditions imported from the above works. The simulations exhibit strong agreement with the experimental data. It also demonstrates the capabilities to capture various effects such as temperature dependence and external magnetic field.}
\label{Fig:nbn_results}
\vspace{-10pt}
\end{figure*}

For NbN, we follow the vortex crossing approach to compute the SNSPD's dark count rate and system detection efficiency. Based on the quasi-particle diffusion model, we calculate the free energy barrier of vortex crossing $U_{max}$ and apply it to our temperature-dependent vortex-crossing model. An initial modeling of the vortex attempt rate $\alpha_v(T)$ is performed based on the dark count data in the work of Zhang \textit{et al.}\cite{zhang_nbn_snspd} for NbN as shown in Fig. \ref{Fig:nbn_results}(a) and (d). We found that $\alpha_0$ is 574kHz and $\beta_1$ is 0.175 meV. 

A list of parameters that have been used in our temperature-dependent vortex crossing model is shown below in Table \ref{tab:table1}. The material parameters are based on the experimental results in works of Zhang \textit{et al.}\cite{zhang_nbn_snspd} and Charaev \textit{et al.}\cite{charaev_snspd}. Based on those parameters, we are able to predict dark count rate and the plateau region of SDE, which align closely with experimental measurements as demonstrated in Fig. \ref{Fig:nbn_results}. It is worth noting that we use the same $\alpha_0$ and $\beta_1$ to predict the dark count rate in Charaev's work, even though their designs are completely different. Comparing the results in Fig. \ref{Fig:nbn_results}(c) and (f), we demonstrate the material-dependent property of the vortex attempt rate. The parameters $\alpha_0$ and $\beta_1$ which define the temperature dependence of the attempt rate, can remain consistent across different device geometries fabricated from the same superconducting material, suggesting their potential transferability between systems.

\begin{table}[h]
\centering
\scriptsize
\renewcommand{\arraystretch}{0.5}
\begin{tabular}{l | l | l }
Parameters & Values for Fig. \ref{Fig:nbn_results}(d,e) & Values for Fig. \ref{Fig:nbn_results}(f)\\
\hline \hline
Bias Current & 0.4$\sim$30 \(\mu A\) & 46 \(\mu A\) \\
Width & 80 nm & 89 nm \\
Thickness & 7.5$\sim$8 nm & 5 nm \\
\(T_c\) & 7.2$\sim$8.1 K & 4.2 K\\
Band gap \(\Delta\) & 1.27$\sim$1.47 meV & 2.47 meV \\
\(\lambda\) & 473$\sim$564 nm & 289 nm \\
De & 36$\sim$44 \(nm^2/ps\) & 54 \(nm^2/ps\) \\
\(\xi\) & 4.5$\sim$5.1 nm & 4.10 nm \\
\(\alpha_0\) & 574 kHz & 574 kHz \\
\(\beta_1\) & 0.175 meV & 0.175 meV \\
\end{tabular}
\caption{NbN Simulation Parameters}
\label{tab:table1}
\end{table}

Notably, an external magnetic field is applied to the SNSPD system in the experiment of Fig. \ref{Fig:nbn_results}(c). To capture the effect of the applied magnetic field $H$, we incorporate an additional term\cite{bulaevskii_vortex-assisted_2012} in the London model to calculate the barrier energy with a uniform field as
\begin{eqnarray}
    \frac{U_{max}}{\epsilon_0} =\nonumber && \max_{x_v}\left[\ln{\left(\frac{2W}{\pi\xi}\sin{(\pi x_v)}\right)}\right.\\&&\left. - \frac{I}{I_c}\frac{2x_v}{\exp(1)\xi} + \frac{H}{H_0}x_v\left(1-\frac{x_v}{\pi}\right)\right]
    \label{eq:potential_function}
\end{eqnarray}
where $x_v = x/W$ is the normalized vortex position, $I_c$ is the critical current at bias temperature $T$, and $H_0=\Phi_0/2W^2$. The model effectively captures temperature-dependent and magnetic field variations in detector performance, demonstrating its applicability across a range of operating conditions. This capability enables reliable predictions of SNSPD responses under different environments. 

\begin{figure*}[htb!]
\centering
\includegraphics[width=\textwidth]{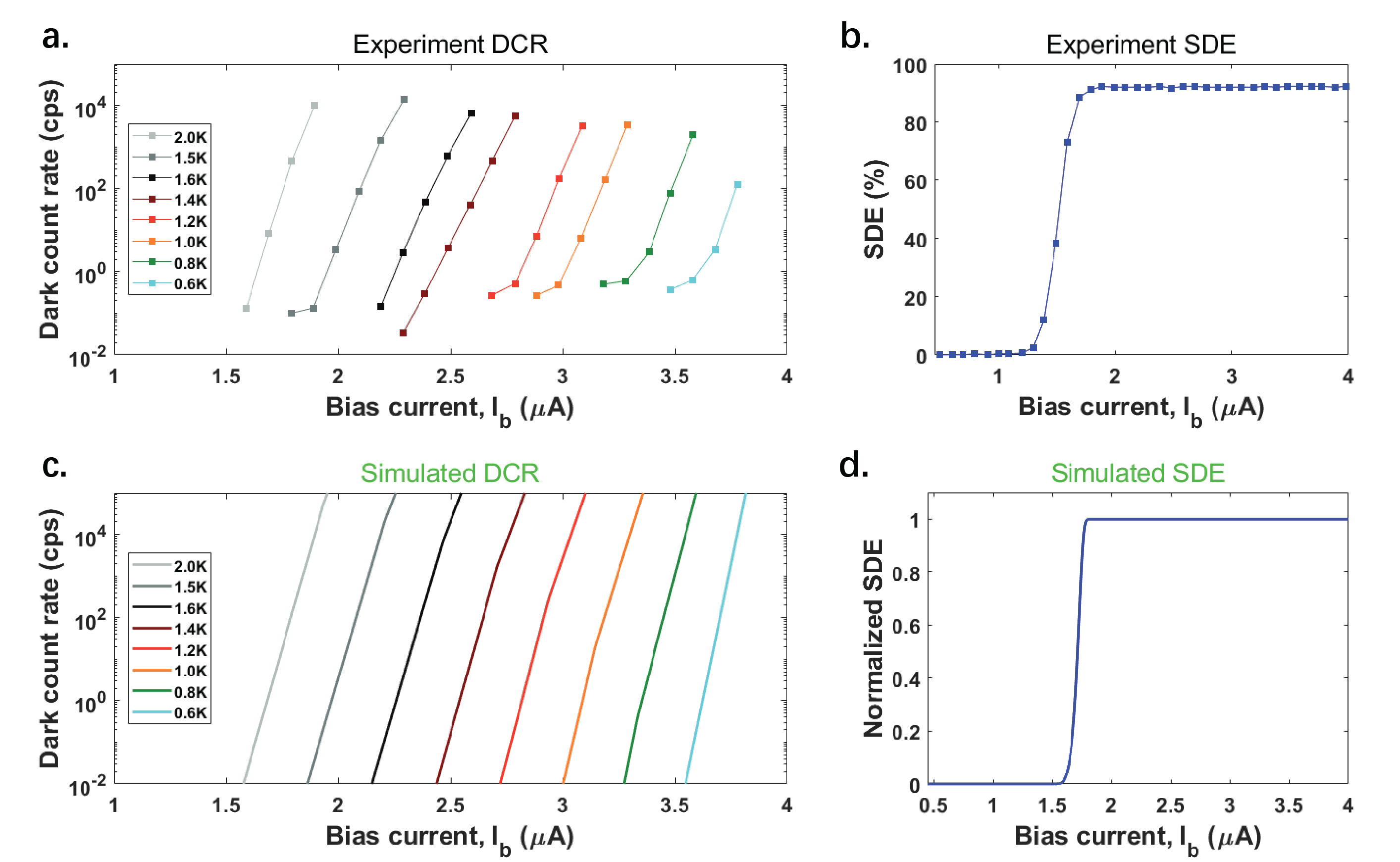}
\caption{WSi based SNSPD Experiment Vs. Simulation. The first row (a)\&(b) shows the experimental results of measured dark count rate (DCR) and system detection efficiency (SDE) retrieved from (a)\&(b) Marsili's work \cite{marsili_detecting_2013}. The second row (c-d) shows the simulation results of our vortex crossing model based on the material parameters and test conditions imported from the above works. Close agreement is found, especially at higher
temperatures.}
\label{Fig:wsi_results}
\vspace{-5pt}
\end{figure*}

\subsection{WSi based SNSPD}
Building upon our analysis of NbN-based SNSPDs, we also perform extensive studies on WSi, a superconductor that also offers distinct advantages in single-photon detection\cite{marsili_detecting_2013}. We calculate the system detection efficiency and dark count rate for a WSi SNSPD using material parameters from Marsili \textit{et al.} \cite{marsili_detecting_2013}. A list of the parameters used in the vortex simulation are provided in Table \ref{tab:table3}. We use the results from Marsili \textit{et al.}\cite{marsili_detecting_2013} of dark count vs. bias current at different temperatures to calibrate the temperature dependent model. In Fig. \ref{Fig:wsi_results}(a)\&(c), we compare simulated and experimental results of the dark count rate vs. bias current under various temperatures. In Fig. \ref{Fig:wsi_results}(b)\&(d), we compare simulated and experimental results of the bias dependent system detection efficiency. We find close agreement in both cases.

\begin{table}[h]
\centering
\scriptsize
\renewcommand{\arraystretch}{0.5}
\begin{tabular}{l | l}
Parameters & Values for Fig. \ref{Fig:wsi_results} \\
\hline \hline
Bias Current & 0.4$\sim$4 \(\mu A\) \\
Bias Temperature & 0.4$\sim$2 K \\
Width & 120 nm \\
Thickness & 4.5 nm\\
\(T_c\) & 3 K \\
Band gap $\Delta$ & 0.59 meV \\
\(\lambda\) & 768 nm\\
De & 71 \(nm^2/ps\)\\
\(\xi(0)\) & 7 nm\\
\(\alpha_0\) & 124.3 MHz\\
\(\beta_1\) & 0.138 meV \\
\end{tabular}
\caption{WSi Simulation Parameters}
\label{tab:table3}
\vspace{-10pt}
\end{table}

\subsection{Timing Jitter}

Timing jitter in SNSPDs can be broadly categorized into two types: extrinsic and intrinsic. Extrinsic timing jitter arises mainly from electronic noise introduced by circuit components such as amplifiers, ground loops, oscilloscopes, and other instrumentation \cite{You_jitter,zhao2011intrinsic}. These sources of noise induce fluctuations in the signal pulses generated by the SNSPD, leading to variability in the time required for the signal to reach the detection threshold. Since extrinsic jitter is highly dependent on the specific characteristics of the experimental setup and instrumentation, its magnitude can vary significantly between different systems. In this work, we focus on the intrinsic timing jitter of SNSPDs, which refers to the uncertainty in the time delay between the arrival of a photon at the detector and the generation of a detection process signal. This timing jitter primarily comprises the longitudinal geometric jitter and the vortex-induced jitter. 

\begin{figure*}[ht!]
\centering
\includegraphics[width=\textwidth]{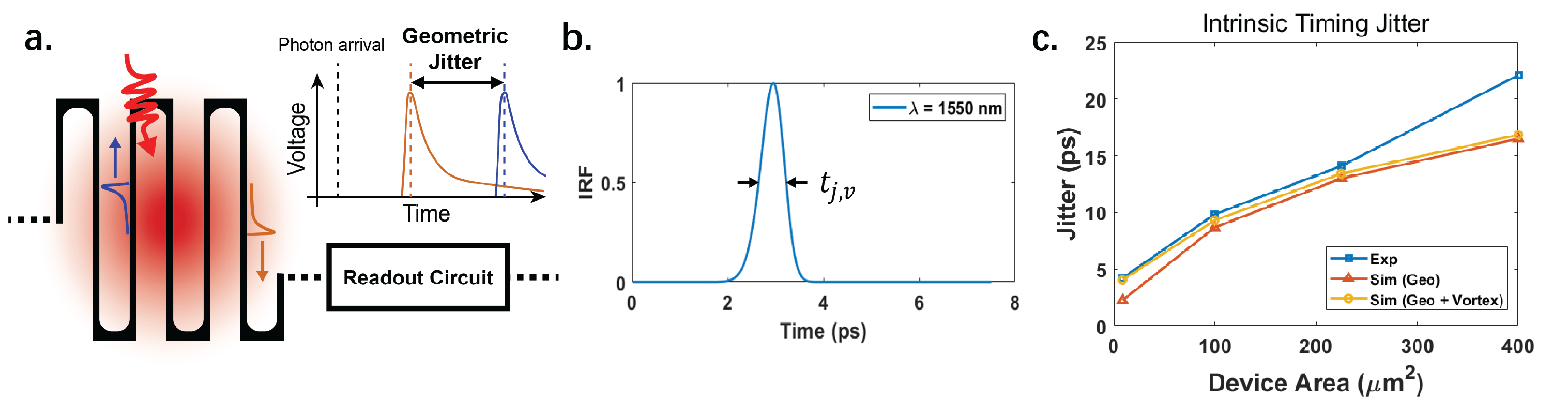}
\caption{The origin of intrinsic timing jitter in SNSPD. (a) Longitudinal geometric contribution to timing jitter. Differences in the distance microwave signals travel along the length of the SNSPD lead to different delay times in the onset of the detection pulse. (b) Instrument response function (IRF) for photons at a wavelength of 1550 nm. The vortex-induced timing jitter $t_{j,v}$ is calculated from the FWHM of the IRF distribution.  (c) Timing jitter vs. device area ranges from $9\mu m^2$ to $400\mu m^2$. The experimental data (blue) are gathered from Calandri's work \cite{calandri_superconducting_2016}, the orange line with triangle markers represents the simulated geometric jitter, and the yellow line with circle markers represents the simulated total jitter.}
\label{Fig:timingjitter}
\vspace{-5pt}
\end{figure*}

For a sufficient detection area, most SNSPDs are designed with meandered superconducting nanowires. Recent studies \cite{calandri_superconducting_2016, 10.1063/1.4960533, 10.1063/1.4954068} have pointed out that these nanowires (usually above 10 $\mu m$) are acting like transmission lines with large inductivity. The speed of the signal transmission in these lines is slower than the speed of light. For an NbN nanowire in this dimension, the transmission speed is reported to be about 2$\%$ of the speed of light \cite{zhao_single-photon_2017}. Such discovery has suggested a longitudinal geometric contribution to the intrinsic timing jitter. As shown in Fig. \ref{Fig:timingjitter} (a), the hot spot induced by photon absorption activates a certain portion of the SNSPD active area, generating signal pulses at different locations. Those signals travel along the length of the SNSPD at different distances, leading to different delay times in the onset of the detection pulse. The difference between the pulse delays is defined as the geometric jitter. This geometric jitter profile also depends on the geometric setup of the nanowire (meandered / straight wires), and can be compensated or minimized by using a short straight wire \cite{korzh_demonstration_2020}.

Here we demonstrate a simplified longitudinal geometric jitter model to estimate the geometric timing jitter for SNSPDs with meandered nanowire design. In Fig. \ref{Fig:timingjitter} (a), we provide an example of how the geometric jitter is being calculated based on the signal generation distance to the readout port. In this case, we assume the signal pulse propagates in one direction (one-end readout). Based on the parameters for geometry setup, including active area size, nanowire width, and pitch, the distance $D(x,y)$ toward the readout port from the arbitrary location of the nanowire can be calculated. The time for the signal to reach the readout port depends on the signal propagation speed $v$  and distance $D(x,y)$ in the nanowire. The distribution of photons depends on the mode profile of the incident light. In our model, we have adopted the Monte-Carlo method to simulate triggered locations inside the spot area with a Gaussian distribution, assuming the mean location is at the center of the active area with a standard deviation of $1\mu m$. The latency calculated in $N$ attempts is defined as:
 \begin{equation}
     t_{l,geo,N}(x,y)=\frac{D_N(x,y)}{v}
     \label{eq:lat_geo}
 \end{equation}
The average latency is given by the mean value $t_{l,geo} = \overline{t_{l,geo,N}}$, and the corresponding geometric timing jitter is represented as the standard deviation of the geometric latency, $t_{j,geo}=std(t_{l,geo,N})$.

As mentioned earlier, both the extrinsic jitter and the geometric jitter can be minimized or even eliminated by using low-noise cryogenic equipment on a short enough ($\sim5\mu m$) superconducting nanowire. In this case, there is still a remnant timing jitter being measured in the system \cite{korzh_demonstration_2020,allmaras_modeling_2020}. These small jitters are dominated by the detection mechanisms of the SNSPD. In our temperature-dependent vortex crossing model which we have proposed and demonstrated previously, we have analyzed this jitter to be induced by the vortex crossing behavior. 
Recalling from the previous section, in our model, the time-dependent vortex crossing rate can be defined as
\begin{equation}
     \Gamma_v(t)=\alpha_v(T)\exp[-U_{max}(t)/k_BT]
     \label{eq:crossing_rate}
 \end{equation}
The probability of detecting a single photon $P_1$ in the time interval $(0,t)$ can be represented as\cite{jahani_probabilistic_2020}
\begin{equation}
     P_1(t)=1-P_0(0,t)=1-\exp\left[-\int^t_{t_0=0}\Gamma_v(t')dt'\right]
     \label{eq:detection_probability}
 \end{equation}
where $P_0$ is the probability that zero counts occurred. In SNSPDs the instrument response function (IRF) defines the temporal distribution of detection events (e.g. vortex crossing) at time $t$ and is represented as the time derivative of $P_1$:\cite{jahani_probabilistic_2020}
\begin{equation}
     IRF(t)=\frac{dP_1(t)}{dt}
     \label{eq:IRF}
 \end{equation}
As shown in Fig. \ref{Fig:timingjitter} (b), the average latency $t_{l,v}$ of such a detection event induced by vortex crossing is defined as the average of the IRF, and the timing jitter $t_{j,v}$ is represented as the FWHM of the IRF distribution. Therefore, the average latency and intrinsic timing jitter are calculated as 
\begin{equation}
     t_{l}= t_{l,geo}+t_{l,v}
     \label{eq:lat_tot}
 \end{equation}
 \begin{equation}
     t_{j}=\sqrt{t_{j,geo}^2+t_{j,v}^2}
     \label{eq:jitter_tot}
 \end{equation}

We compared our simulation results with the experimental data reported by Calandri et al.,\cite{calandri_superconducting_2016}, who investigated the timing jitter of NbN-based SNSPDs as a function of device area, as shown in Fig. \ref{Fig:timingjitter}. In their study, the extrinsic timing jitter was minimized through careful experimental design, allowing the measured jitter to predominantly reflect intrinsic contributions. The examined device areas ranged from $9\mu m^2$ to $400\mu m^2$, with a nanowire width of $100 nm$ and a pitch of $200 nm$. Our model shows strong agreement with the experimental data for device areas up to approximately $300\mu m^2$. A noticeable deviation occurs at $400\mu m^2$, which may be attributed to variations in incident photon mode profile across larger devices. Despite this discrepancy, both our simulations and the experimental results consistently indicate that intrinsic timing jitter is primarily governed by the geometric properties of the SNSPDs.

We have demonstrated that the vortex crossing model can be used to predict SNSPD behavior in NbN and WSi. In NbN we demonstrate close matching with experiments of the temperature dependent dark count rate, magnetic field dependent dark count rate, and plateau region of SDE. Our model also shows close matching with WSi experiments of temperature dependent dark count rate and SDE plateau region. Lastly, we present a model for the origin of intrinsic timing jitter in SNSPDs which matches closely with experiments over a range of device areas. Therefore, our model captures many of the important device metrics for recent SNSPD applications.

\begin{acknowledgments}
This work was funded by the DARPA SynQuaNon program. We would like to thank Prasanna V. Balachandran and Shunshun Liu for their fruitful discussions on superconducting material properties.
\end{acknowledgments}

\section*{Data availability statement}
The data that support the findings of this study are available from the corresponding author upon reasonable request.


%
%

%


\bibliography{refs}

\end{document}